\documentclass[fleqn,usenatbib]{mnras}

\usepackage{newtxtext,newtxmath}

\usepackage[T1]{fontenc}
\usepackage{ae,aecompl}
\usepackage{ulem}

\usepackage{graphicx}	
\usepackage{amsmath}	

\def\nu{\textit{NuSTAR}}

\def\int{\textit{INTEGRAL}}
\def\heao{\textit{HEAO-1}}
\def\rxte{\textit{RXTE}}
\def\sci{\textit{SCI}}
\def\occ{\textit{OCC}}

\def\asci{a\textit{SCI}}
\def\aocc{a\textit{OCC}}

\def\ergscm{erg~s$^{-1}$ cm$^{-2}$ deg$^{-2}$}
\def\flux{erg s$^{-1}$ cm$^{-2}$}

\title[NuSTAR CXB measurement]{NuSTAR measurement of the cosmic X-ray background in the 3--20 keV energy band}

\author[Krivonos et al.]{Roman Krivonos,$^{1}$\thanks{E-mail: krivonos@cosmos.ru (IKI)}
Daniel Wik,$^{2}$ 
Brian Grefenstette,$^{3}$
Kristin Madsen,$^{4}$ \newauthor
Kerstin Perez,$^{5}$ 
Steven Rossland,$^{2}$ 
Sergey Sazonov,$^{1,6}$ 
and Andreas Zoglauer$^{7}$\\
$^{1}$Space Research Institute, Russian Academy of Sciences, Profsoyuznaya 84/32, 117997 Moscow, Russia\\
$^{2}$Department of Physics and Astronomy, University of Utah, Salt Lake City, UT 84112, USA;\\
$^{3}$Space Radiation Laboratory, Caltech, 1200 E California Blvd, Pasadena, CA 91125, USA;\\
$^{4}$CRESST and X-ray Astrophysics Laboratory, NASA Goddard Space Flight Center, Greenbelt, MD 20771, USA\\
$^{5}$Department of Physics, Massachusetts Institute of Technology, Cambridge, MA 02139, USA;\\
$^{6}$Moscow Institute of Physics and Technology, Dolgoprudny, Moscow Region, 141700, Russia;\\
$^{7}$Space Sciences Laboratory, University of California, Berkeley, CA 94720, USA;
}

\date{Accepted XXX. Received YYY; in original form ZZZ}

\pubyear{2020}

\begin{document}
\label{firstpage}
\pagerange{\pageref{firstpage}--\pageref{lastpage}}
\maketitle

\begin{abstract}
We present measurements of the intensity of the Cosmic X-ray Background (CXB) with the \nu\  telescope in the 3--20~keV energy range. Our method uses spatial modulation of the CXB signal on the \nu\ detectors through the telescope's side aperture. Based on the \nu\ observations of selected extragalactic fields with a total exposure of 7~Ms, we have estimated the CXB 3--20~keV flux to be $2.8\times 10^{-11}$~\ergscm, which is $\sim8\%$ higher than measured with \heao\ and consistent with the \int\ measurement. The inferred CXB spectral shape in the 3--20~keV energy band is consistent with the canonical model of Gruber et al. We demonstrate that the spatially modulated CXB signal measured by \nu\ is not contaminated by systematic noise and is limited by photon statistics. The measured relative scatter of the CXB intensity between different sky directions is compatible with cosmic variance, which opens new possibilities for studying CXB anisotropy over the whole sky with \nu.
\end{abstract}

\begin{keywords}
(cosmology:) cosmic background radiation -- galaxies: active -- instrumentation: detectors -- X-rays: general -- methods: data analysis
\end{keywords}



\section{Introduction}

The cosmic X-ray background (CXB) has been studied by virtually all X-ray observatories since its discovery in 1962 by \cite{1962PhRvL...9..439G}. At low energies (below 10~keV), the bulk of the CXB has been directly resolved into discrete X-ray sources, primarily active galactic nuclei (AGNs, e.g. \citealt{2017ApJS..228....2L}), and it is widely believed that the CXB is mostly composed of AGNs also at higher energies. In fact, the Nuclear Spectroscopic Telescope Array \citep[\nu,][]{2013ApJ...770..103H}, thanks to its focusing hard X-ray optics, has recently resolved 33\%--39\% of the CXB in the 8--24~keV energy band \citep{2016ApJ...831..185H}, i.e. close to the peak of the CXB spectrum (at $\sim 30$~keV), by measuring AGN number counts in deep extragalactic fields. 

\begin{table*}
\caption{List of the \nu\ deep extragalactic observations analyzed in this work.}
\label{tab:data}
\begin{tabular}{cccccc}
\hline
Field & R.A. & DEC. & Area (sq. deg) & Raw exposure & Ref.\\
\hline
COSMOS & 150.2 & 2.2  & 1.7 & 3.1 Ms & \cite{2015ApJ...808..185C} \\
EGS & 214.8 & 52.8 & 0.18 & 1.6 Ms & \cite{2007ApJ...660L...1D} \\
ECDFS & 53.1  & -27.8 & 0.25 & 1.5 Ms & \cite{2015ApJ...808..184M} \\
UDS & 34.4 & -5.1 & 0.4 & 1.7 Ms & \cite{2018ApJS..235...17M} \\
\hline
\end{tabular}
\end{table*}

There is great benefit in measuring the intensity and broad-band spectrum of the CXB as precisely as possible since that would provide strong constraints on the evolution of the AGN population with cosmic time and its composition (in particular, the relative fractions of obscured, unobscured, and beamed sources), complementary to the information inferred from the statistics of AGNs provided by deep X-ray surveys (see e.g. \citealt{2014ApJ...786..104U}). Moreover, the CXB is expected to contain significant contributions of other classes of extragalactic objects, such as X-ray binaries in normal and starburst galaxies (e.g. \citealt{2012MNRAS.421..213D,2016ApJ...825....7L}) and even the seeds of supermassive black holes at Cosmic Dawn (e.g. \citealt{2018MNRAS.481.3278R}).  

Apart from global properties of the CXB, there is interest in studying its variations over the sky, which should carry information on the large-scale structure on the Universe \citep[e.g.,][]{1983IAUS..104..333S}. The expected anisotropy at large angular scales is at a few per cent level, which presents a big challenge for most X-ray experiments. Nevertheless, significant angular variations in the CXB intensity have been unveiled in the \heao\ data \citep[e.g.,][]{1991ApJ...378L..37J,1994ApJ...434..424M} and more confidently in the \rxte\ data \citep{2008A&A...483..425R}. The latter measurement made it possible to estimate the integrated X-ray emissivity of the local volume of the Universe (within $\sim 100$~Mpc) and put important constraints on the populations of relatively faint X-ray sources such as low-luminosity AGNs. Nevertheless, there is a strong need in verification and improvement of these measurements. 

CXB measurements in the energy range from tens to hundreds keV are complicated because the signal recorded by the detector is often dominated by the internal background, caused by the interactions of charged particles with the detector material and spacecraft structures. Since CXB radiation is coming from all directions in the sky, i.e. it pervades the full field of view (FOV) of X-ray telescopes, it is impossible to estimate the local background as in the case of point-like or slightly extended X-ray sources. The question of dissecting the observed total X-ray background into particle-induced background and a CXB contribution is often addressed by modelling the internal detector background or by modulating the relative contributions of these two components. The former approach requires that one has a very good model of the internal background. The latter one is often more effective, because modulating methods are usually carried out by geometric design, and as a result, better determined.

The 3--80~keV working energy range of \nu\ makes it possible to study the CXB at its intensity peak. On the other hand, the open geometry of the mast that separates the \nu\ X-ray optics from the detectors, allows any X-ray radiation, including the CXB, to shine directly on the detectors \citep{2017arXiv171102719M}. This issue, known as stray-light or aperture flux (the light not focused by the optics) greatly complicates the observations of X-ray sources \citep[e.g.,][]{2014ApJ...781..107K,2015ApJ...815..132Z,2015ApJ...814...94M,2017ApJS..229...33F,2017PhRvD..95l3002P,2019ApJ...877...96T}, but also opens a possibility to study the CXB integrated emission. 

As mentioned above, both detector background modelling and signal modulation can be used for CXB measurements. \cite{2014ApJ...792...48W} has developed a sophisticated spectral background model for the \nu\ detectors, which has been successfully used in many relevant studies  \citep[e.g.,][]{2014ApJ...791...81A,2015ApJ...800..139G,2016ApJ...831..185H,2018NatAs...2..731H}. However, the usage of the \nu\ background model for CXB measurements is complicated by many factors like overabundance of model parameters (more than a hundred for one detector, mainly needed for modelling a large number of instrumental emission lines), periodic activation by South Atlantic Anomaly (SAA) passages, poorly investigated and variable emission component from the Sun (and potentially Earth's albedo), long-term variations due to energetic particles in the radiation belts during periods of high solar activity \citep{radbelts}, etc. 

The purpose of this study is to measure the CXB intensity using data from the \nu\ extragalactic survey program, taking advantage of deep observations of selected extragalactic fields. To overcome the aforementioned difficulties, we take a CXB modulation approach and use the fact that the CXB produces a well-determined spatial gradient on the \nu\ detectors. We treat all instrumental uncertainties as a single uniform detector component in contrast to the CXB spatially modulated background component. This allows us to make robust measurements of the CXB intensity in spectral bins up to 20~keV. At higher energies, \nu\ background modelling becomes complicated, because instrumental background starts to dominate \citep{2014ApJ...792...48W} and the contribution of Earth's hard X-ray emission, associated with the reflection of the CXB and cosmic ray interactions, becomes significant \citep{2007MNRAS.377.1726S,2008MNRAS.385..719C}.

\section{Observations and data processing}

We use data from the \nu\ extragalactic survey program, which includes a number of well-known fields (Table~\ref{tab:data}) with different sky coverage and depth: a shallow, wide-area survey of the COSMic Evolutionary Survey field \citep[COSMOS,][]{2015ApJ...808..185C}, a deep, pencil-beam survey of the Extended Chandra Deep Field-South \citep[ECDFS,][]{2015ApJ...808..184M}, observations of the Extended Groth Strip \citep[EGS,][]{2007ApJ...660L...1D}, and of the UKIDSS Ultra Deep Survey \citep[UDS,][]{2018ApJS..235...17M}.

\nu\ has two identical co-aligned telescopes, each consisting of an independent set of X-ray mirrors and a focal-plane detector, referred to as focal plane module (FPM) A and B (FPMA and FPMB). The optics are based on the grazing incidence conical approximation Wolter~I design, where incoming X-ray photons are focused by reflection from an upper and lower cones. The FOV for these ``two-bounce'' photons, determined by the detector dimensions,  is $\sim13' \times 13'$. Each focal plane module is composed of four $32\times32$ solid state pixel detector arrays (or ``chips'', referred to as DET0, DET1, DET2, and DET3). Each detector pixel has a size of 0.6~mm. The \nu\ CdZnTe sensors provide a minimum detector threshold of 1.6--2~keV and ability to register X-ray photons up to $\sim160$~keV \citep{2009SPIE.7435E..03R,2011SPIE.8145E..07K,brian2017}. Note that the \nu\ mirrors' collecting area is limited by the PT K-edge at 78.4 keV. The response across a given FPM detector is largely uniform. However, during ground-based characterization, \cite{2009SPIE.7435E..03R} reported on non-uniformity of the detector pixel arrays at a level larger than statistical fluctuations \citep[see also][]{2012PhDT.........5B}. Based on our experience with in-flight \nu\ data, the relative per-chip internal background count rate can be different by 5--10\% with respect to the mean level (except FPMA/DET3 with a $\sim$20\% deviation), probably due to the different thickness of the chips.

\begin{figure*}
  \includegraphics[width=0.99\textwidth]{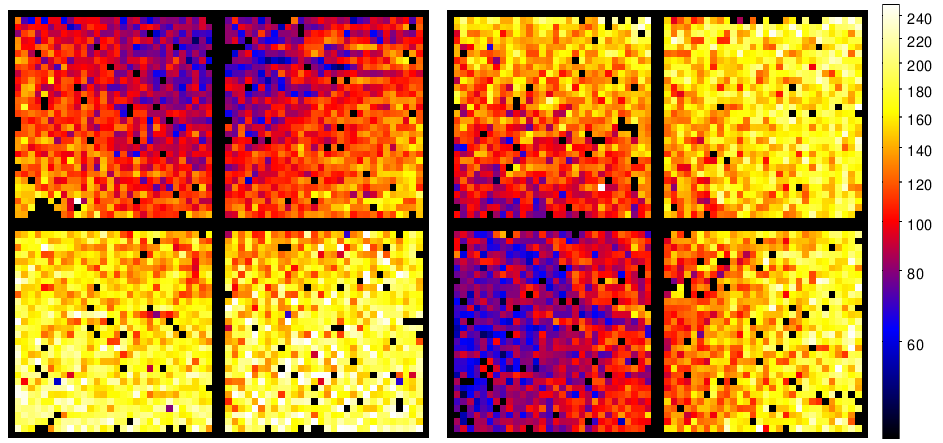}
\caption
   { \label{fig:cosmos:stack} 
\nu\ FPMA (left) and FPMB (right) 3--20~keV stacked images of the COSMOS field in detector coordinates. The image demonstrates the total counts registered in each detector pixel during the exposure time of 2.4~Ms. The square root color map ranges from 50 to 250 counts per pixel.}
\end{figure*}

Initial data reduction was performed with the \nu\ Data Analysis Software pipeline ({\it NuSTARDAS}) v1.8.0. To reduce background uncertainties, we used the general \nu\ data processing routine {\it nupipeline} with flags SAAMODE=STRICT and TENTACLE=yes, which removed all data from passages through the South Atlantic Anomaly (SAA). In addition to the normal \nu\ observing scientific mode (hereafter \sci), when aspect solution is available from the on-board star tracker located on the X-ray optics bench (Camera Head Unit $\#$4, CHU4; see e.g., 
{\it The NuSTAR Data Analysis Software Guide}\footnote{\url{https://heasarc.gsfc.nasa.gov/docs/nustar/analysis/nustar_swguide.pdf}} for details), i.e. when CHU4 is not shielded by Earth, we also extracted cleaned events in occultation mode, when the FOV is blocked by Earth (hereafter \occ). Note that the optical axis is always pointed at the target position, even during Earth occultation periods.

We then checked the light curves for FPMA and FPMB produced from the cleaned events, and excluded observations with a 3--10~keV count rate higher than 0.17 counts s$^{-1}$, as a tracer of increased background due to increased solar activity. Using this condition, we reduced the data set for each extragalactic field as follows. From the list of 125 COSMOS observations, we removed 28 (003, 019, 020, 021, 022, 055, 056, 057, 058, 067, 076, 081, 083, 084, 087, 088, 089, 090, 091, 092, 093, 096, 097, 110, 111, 117, 119, 120), leaving 97 observations with a total cleaned exposure time of 2.4~Ms. Two observations (60022011002 and 60022003001) were removed from the ECDFS field, and, additionally, observation 60022001001 was excluded because of low exposure (700~s), leaving 31 clean observations with a total exposure of 1.4~Ms. For the EGS field with 32 observations in total, we excluded 60023007005 and 60023008003, leaving 30 observations with a total exposure of 1.5~Ms. The UDS field has no observations exceeding the filtering threshold, so that all 35 observations with a total exposure of 1.7~Ms were used in the analysis.

The COSMOS field has the deepest exposure and longest duration in time, from Dec 2012 to Feb 2014. To minimize possible influence of long-term background variations on the analysis, we divided the COSMOS observations into three epochs: COSMOS {\it EP1} from 26 Dec 2012 to 20 Jan 2013 (750~ks); COSMOS {\it EP2} from 3 Apr 2013 to 21 May 2013 (630~ks), and COSMOS {\it EP3} from 3 Dec 2013 to 25 Feb 2014 (1020~ks). The resulting 6 clean data sets used in the subsequent analysis are listed in Table~\ref{tab:data:num}, where also their numbering from 1 to 6 is introduced. 

\begin{table}
\caption{Data sets used in the analysis.}
\label{tab:data:num}
\begin{tabular}{lccccc}
\hline
ID & Field & Begin & End &  $T_{\rm exp}$ \\
\hline
1 & COSMOS {\it EP1} &  26-12-2012 & 20-01-2013 & 750~ks  \\
2 & COSMOS {\it EP2} &  03-04-2013 & 21-05-2013 & 630~ks  \\
3 & COSMOS {\it EP3} &  03-12-2013 & 25-02-2014 & 1020~ks \\
4 & EGS & 15-11-2013 & 27-11-2014 & 1.5~Ms \\
5 & ECDFS & 28-09-2012  & 01-04-2013 & 1.4~Ms \\
6 & UDS & 24-01-2016 &  18-11-2016 & 1.7~Ms \\
\hline
\end{tabular}
\end{table}

Instead of building telescope images in sky coordinates, as is usually done in treating focused \nu\ observations, in this work we use the detector coordinate system, which is native for stray-light observations \citep{2017ApJ...841...56M}. In the following, we consider each \nu\ image in RAW detector coordinates as a default \citep[however, in contrast to][who use DET1]{2017ApJ...841...56M}, unless otherwise stated. Note that stray-light studies can be carried out both in DET1 and RAW coordinates. The former naturally account for spatial non-uniformities in the pixel response. However, we found that our fitting procedure is more stable in bigger RAW pixels, as having more counts, which is especially important in narrow energy bands. For this reason we used RAW detector pixels taking, at the same time, the calibrated pixel response into account. We generated FPMA and FPMB images by  combining sub-detector $32\times32$ coordinates (RAWX, RAWY) into $64\times64$ arrays. The images were generated in 20 energy intervals logarithmically spaced between 3 and 20 keV. Note that we did not subtract the contribution from detected point sources, treating them as part of the CXB. However, we checked that their total flux is negligible compared to the observed detector count rate. We should also note that the detected sources are effectively averaged in detector coordinates, and do not produce any significant spatial variations.


\section{CXB model}

The canonical CXB spectrum is based on the \heao\ data parameterized by \cite{1999ApJ...520..124G} in a broad energy range.
We adopt the analytic approximation for the 3--60~keV range from \cite{1999ApJ...520..124G}, as it covers the \nu\ energy band (3--20~keV) studied in this paper:
\begin{equation}
\label{eq:g99}
S_{\rm CXB}^{\rm G99} (E)=7.877 E^{-0.29} e^{-E/41.13}.
\end{equation}
We have rescaled $S_{\rm CXB}^{\rm G99}$, originally given by \cite{1999ApJ...520..124G} in units of keV/keV~cm$^{-2}$~s$^{-1}$~sr$^{-1}$, to keV/keV~cm$^{-2}$~s$^{-1}$~deg$^{-2}$ as more convenient for our analysis, and constructed the corresponding {\sc xspec} model for spectral fitting.

\cite{2007A&A...467..529C} reported on a 10\% higher CXB normalization using special Earth occultation observations by \int\ \citep[see also ][]{2010A&A...512A..49T}. We use both \cite{1999ApJ...520..124G} and \cite{2007A&A...467..529C} as reference CXB measurements in this work. Additionally, we compare our results with \rxte\ measurements of the CXB, performed in the same energy band of 3--20~keV by \cite{2003A&A...411..329R}.

\section{Detector background model}
\label{sect:model}

The detector background of \nu\ is investigated in the papers \cite{2014ApJ...792...48W,2017arXiv171102719M} and can be summarized as follows. At any given time, the detector count rate of the \nu\  FPMA and FPMB consists of:

\begin{itemize}
\item[--] CXB;  
\item[--] emission from point-like and extended X-ray sources;
\item[--] Galactic Ridge X-ray background, if the telescope's FOV is directed toward the Galactic plane;
\item[--] detector internal background. 
\end{itemize}

Astrophysical sources of X-ray emission (the first three items on the list) can be registered by the \nu\ focal plane detectors in two ways: through the mirror optics, i.e. focused X-rays; and directly from the telescope's sides at 1--4 degrees away from the optical axis, which is usually referred to as stray-light or aperture flux \citep{2017arXiv171102719M}. The detector background count rate in the 3--20~keV band is dominated by the aperture CXB (aCXB), which exceeds the focused CXB (fCXB) by almost an order of magnitude \citep{2014ApJ...792...48W}. The aCXB component is clearly visible on the FPMA and FPMB detectors as strong spatial variations. The reason for the spatial variation which manifests as a ramp, is the uneven shadowing of the optical bench onto the focal plane bench. Figure~\ref{fig:cosmos:stack} shows stacked FPMA and FPMB images based on all cleaned COSMOS \sci\ data with a total exposure of 2.4~Ms in the 3--20 keV energy band. Both FPMA and FPMB images demonstrate strong spatial gradient.

Internal detector background is produced by various processes including activation of different elements of the spacecraft and interaction of the detector material with cosmic rays. The background model is characterised by a number of known emission lines and a continuum parameterized by a broken power-law with $E_{\rm break} = 124$~keV \citep{2014ApJ...792...48W}. There is also a background component at low energies $E<5$~keV, presumably associated with the scattered X-ray emission from the Sun, and modelled by \cite{2014ApJ...792...48W} with a $\sim1$~keV collisionally-ionized plasma \citep[{\sc apec} model in {\sc xspec},][]{Smith_2001}. However, this model provides a poor fit to the observed background spectrum, demonstrating a clear excess in the 5--10~keV energy range, as reported by a number of authors \citep{2019PhRvD..99h3005N,2019ApJ...884..153P,2020PhRvD.101j3011R}, who replaced the {\sc apec} model with a power law. In the current spatial analysis, we first attribute this spectral component to a flat instrumental background, allowing it to manifest itself in the residuals. 

\begin{figure*}
  \includegraphics[width=0.99\textwidth]{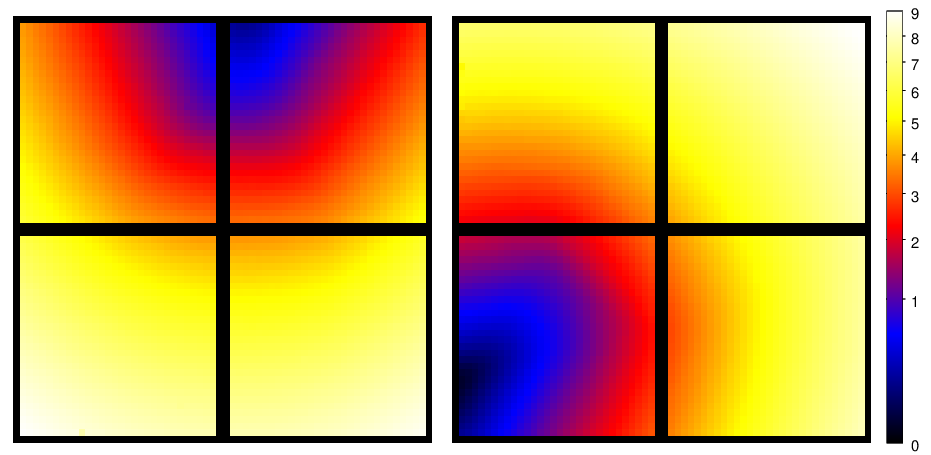}
\caption
   { \label{fig:deg2} 
Image of the \nu\ focal plane modules A (left) and B (right) in physical detector pixels, showing the open portion of the sky in squared degrees. The image also shows detector gaps between detector chips (upper right DET0, upper left DET1, bottom left DET2, and bottom right DET3). The square root color map ranges from 0 to 9 deg$^2$ per pixel.}
\end{figure*}

Each detector pixel is open to the sky at different solid angles, ranging from $\sim0.2$ to $\sim9$ deg$^2$, as shown in Fig.~\ref{fig:deg2}. In addition to varying solid angle, each pixel sees a slightly different portion of the sky with respect to the other pixels; the combined stray-light aperture FOV is shown, e.g., in fig.~9 of \cite{2014ApJ...792...48W}. 

The background model developed in this work contains essentially two basic components:

\begin{itemize}
\item[--] flat detector background. The total number of counts in a given pixel is proportional to a nearly flat instrumental background, which includes internal emission lines and continuum;
\item[--] spatially variable aperture component. The count rate is proportional to the open sky solid angle. \end{itemize}

The main goal of this study is to separate out the flat and aperture components and to provide a physical interpretation for the latter. We naturally expect the aperture component to be dominated by the CXB. Note that, by design of this model, we neglect the focused CXB and contribution of point sources. The focused CXB emission is expected to follow the mirrors' vignetting response. However, it does not manifest itself in deep stacked detector images (Fig.~\ref{fig:cosmos:stack}), which allows us to disregard this component in the analysis. The contribution of detected point sources (mostly AGNs) is also negligible and averaged in detector coordinates.

The method implies maximization of a likelihood function constructed in the following way. The probability $P$ to observe $N_{\rm pix}$ counts in a given pixel is:
\begin{equation}
\label{eq:prob}
P = (N_{\rm bkg}*M_{\rm bkg} + N_{\rm apt}*R_{\rm pix}*A_{\rm det}*A_{\rm Be}*\Box*\Omega)*T,
\end{equation}
where $N_{\rm bkg}$ and $N_{\rm apt}$ are unknowns of the model, which represent the normalization of the flat instrumental and aperture background component, respectively; $M_{\rm bkg}$ describes the non-uniformity of the detector chips and their relative normalization, obtained from \occ\ data in the 10--20~keV energy band; $R_{\rm pix}$\footnote{This term is not needed when stray-light study is carried out in DET1 coordinates.} accounts for relative spatial non-uniformities in the pixel response stored in the \nu\ CALDB; the aperture component is subject to the energy-dependent absorption in the path of the stray light: detector CdZnTe dead layer absorption ($A_{\rm det}$), directly available for each detector chip from the \nu\ CALDB, and Beryllium window efficiency in front of the detectors, $A_{\rm Be}$, which drops sharply below 10~keV; $\Box=0.0036$~cm$^2$ denotes the geometric area of the pixel; $\Omega$ represents the aperture FOV of a given pixel in deg$^2$; $T$ is the total exposure of the individual observation in seconds. Thus, $N_{\rm bkg}$ and $N_{\rm apt}$ have units of cnts~s$^{-1}$~pix$^{-1}$ and cnts~s$^{-1}$~cm$^{-2}$~deg$^{-2}$, respectively. 

We use a log-likelihood function to derive the maximum likelihood estimator of the parameters $N_{\rm bkg}$ and $N_{\rm apt}$:
\begin{equation}
\label{eq:likelihood}
L = -2 \sum_{\rm i}  N_{\rm pix, i}*\log P - P - \log N_{\rm pix, i}! 
\end{equation}
where $i$ runs over the array of pixels (excluding bad pixels) for all observations of a given extragalactic field. The calculation of the aperture FOV parameter $\Omega$ for each detector pixel (Fig.~\ref{fig:deg2}) is based on the known structure of the telescope and done with the {\sc nuskybgd} code \citep{2014ApJ...792...48W}.

We then estimated the $N_{\rm bkg}$ and $N_{\rm apt}$ parameters by maximizing $L$ using the AMOEBA numerical optimization algorithm. We estimated the errors of the best-fit model parameters by bootstrapping, applying a large number of simulations ($10^4$). This procedure was performed for each of the 20 energy bands from 3 to 20~keV.

\begin{figure}
  \includegraphics[width=0.99\columnwidth]{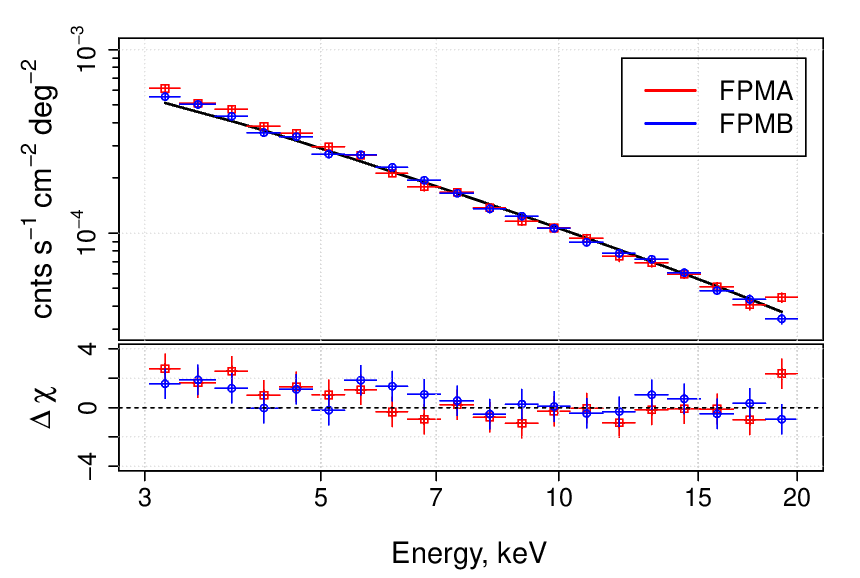}
\caption
   { \label{fig:cosmos:ep1:sci} 
\nu\ spectrum of the aperture background component, obtained from the COSMOS {\it EP1} field using \sci\ data, and its best fit in the 10--20~keV energy range by the CXB model given by equation~(\ref{eq:g99}) with free normalization.}
\end{figure}

Figure~\ref{fig:cosmos:ep1:sci} shows the inferred spectrum of the aperture background component for the COSMOS {\it EP1} field and the result of its fitting in the 10--20~keV energy range by the adopted CXB spectral shape (equation~\ref{eq:g99}). The inferred normalization is $F_{\rm 10-20\,keV}=(1.35\pm0.06)\times 10^{-11}$~\ergscm, with the cross-normalization constant between FPMA and FPMB $C=0.98\pm0.02$ and fit statistics $\chi^{2}_{\rm r}$/dof =$0.80/12$. The estimated flux of the \nu\ aperture background component in the 10--20~keV energy band is $\sim10\%$ higher than that measured by \heao, $F_{\rm 10-20\,keV}=1.23\times 10^{-11}$\ergscm \citep{1999ApJ...520..124G}, but consistent within 1\% with the \int\ measurement, $(1.36\pm0.01\pm0.04)\times 10^{-11}$~\ergscm\ \citep[quoted here are the statistical and systematic errors,][]{2007A&A...467..529C}. This satisfactory agreement suggests that the observed aperture background component is dominated by the CXB. However, the observed excess at soft energies $E<10$~keV (see Fig.~\ref{fig:cosmos:ep1:sci}) clearly disagrees with the CXB spectral shape, which is also revealed by poor fit statistics $\chi^{2}_{\rm r}$/dof =$1.3/38$ in the full 3--20~keV range.

\begin{figure*}
  \includegraphics[width=0.99\textwidth]{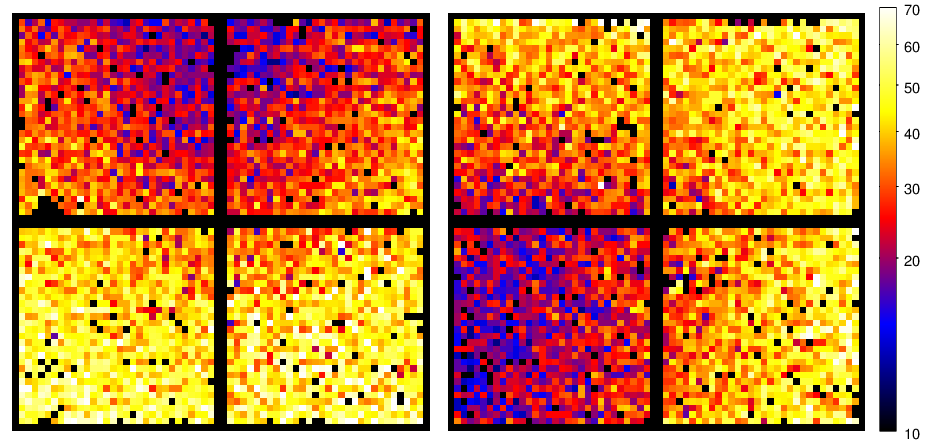}
\caption
   { \label{fig:cosmos:stack:occ} 
\nu\ FPMA (left) and FPMB (right) 3--5~keV stacked images of the COSMOS field in detector coordinates, accumulated during Earth occultation (\occ\ data).  The images demonstrate the total counts registered in each detector pixel during the exposure time of 1.6~Ms. The square root color map ranges from 10 to 70 counts per pixel.}
\end{figure*}

To investigate the spatial distribution of the observed soft excess across the detector, we constructed FPMA and FPMB images of the COSMOS field in the 3--5~keV energy band using the observations when the FOV was blocked by Earth. As seen from Fig.~\ref{fig:cosmos:stack:occ}, the \nu\ \occ\ images demonstrate a significant spatial gradient consistent with that of the CXB in the \sci\ data (hereafter we refer to the aperture components apparent in the \sci\ 3--20~keV and \occ\ 3--5~keV data as \asci\ and \aocc, respectively). This allows us to apply the same background model to the \occ\ data in order to extract spectrum of the \aocc\ component.

\begin{figure}
  \includegraphics[width=0.99\columnwidth]{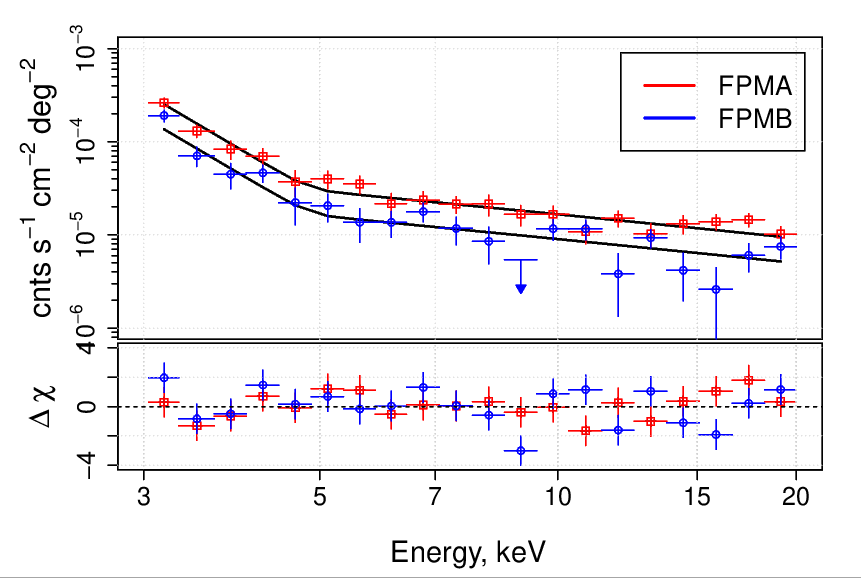}
\caption
   { \label{fig:cosmos:ep1:occ} 
\nu\ spectrum of the aperture background component obtained from the COSMOS {\it EP1} field using Earth occultation (\occ) data, and fitted by a broken power law with $\Gamma_{1}=5.2_{-1.2}^{+1.4}$, $\Gamma_{2}=0.8\pm0.3$, and $E_{\rm break}=4.8_{-0.5}^{+0.9}$~keV.}
\end{figure}

Figure~\ref{fig:cosmos:ep1:occ} shows the spectrum obtained from the COSMOS {\it EP1} \occ\ data with a total exposure of 1.6~Ms. The spectrum demonstrates a significant soft excess and a shallow continuum. The spectrum can be approximated by a broken power law with the following best-fit parameters: $\Gamma_{1}=5\pm1$, $\Gamma_{2}=0.9\pm0.3$, and $E_{\rm br}=4.8\pm0.9$~keV at poor, but acceptable fit statistics $\chi^{2}_{\rm r}$/dof =$1.3/35$. The cross-normalization constant between FPMA and FPMB $C=0.54\pm0.08$ reveals a strong difference between the FPMs. This difference is probably related to the fact that we observe Solar contribution described below, with different normalizations for both FPMA and FPMB based on their relative illumination from the Sun. Note that relative normalization of FPMs is not used in the following analysis and hence, does not affect main results of this work.

We found similar spectral shapes of \aocc\ for all other fields. The corresponding best-fit parameters of the spectral analysis are listed in Table~\ref{tab:bknpower}. The spectral break $E_{\rm br}$ is well fitted at $\sim5$~keV for all the data sets. The COSMOS {\it EP1-3} observations, affected by abnormally high background radiation due to solar flares, demonstrate a steeper soft component ($\Gamma=5-7$) and a higher 3--5~keV flux.

\begin{table}
\caption{Best-fit parameters of the broken power-law model applied to the \occ\ data in the different data sets (Table~\ref{tab:data:num}). The 3--5~keV flux $F_{\rm 3-5\ keV}$ is expressed in units of $10^{-12}$~erg s$^{-1}$ cm$^{-2}$ deg$^{-2}$.}
\label{tab:bknpower}
\begin{tabular}{lccccccccc}
\hline
ID & $\Gamma_{1}$ & $E_{\rm br}$ & $\Gamma_{2}$ & $F_{\rm 3-5\ keV}$ & Const & $\chi^{2}_{\rm r}$/dof \\
   &              & keV          &              &                    &   $10^{-1}$    &               \\
\hline
1 & $5\pm1$ & $4.8\pm0.9$ & $0.9\pm0.3$ & $1.3\pm0.2$ & $5.4\pm0.8$ & 1.3/35 \\ 
2 & $6\pm1$ & $4.6\pm0.6$ & $0.9\pm0.3$ & $1.4\pm0.2$ & $5.5\pm0.7$ & 1.1/35 \\ 
3 & $7\pm1$ & $4.4\pm0.4$ & $1.3\pm0.2$ & $1.8\pm0.1$ & $6.3\pm0.6$ & 1.9/35 \\ 
4 & $4_{-1}^{+2}$ & $5.3\pm1.0$ & $0.9_{-0.7}^{+0.3}$ & $0.8\pm0.1$ & $6.6\pm1.0$ & 1.4/35 \\ 
5 & $4\pm1$ & $5.3\pm0.7$ & $0.8\pm0.2$ & $0.8\pm0.1$ & $6.4\pm0.7$ & 1.1/35 \\ 
6 & $3\pm1$ & $5.1\pm0.7$ & $0.6\pm0.2$ & $0.7\pm0.1$ & $6.2\pm0.6$ & 1.5/35 \\ 
\hline
\end{tabular}
\end{table}

\begin{figure}
  \includegraphics[width=0.99\columnwidth]{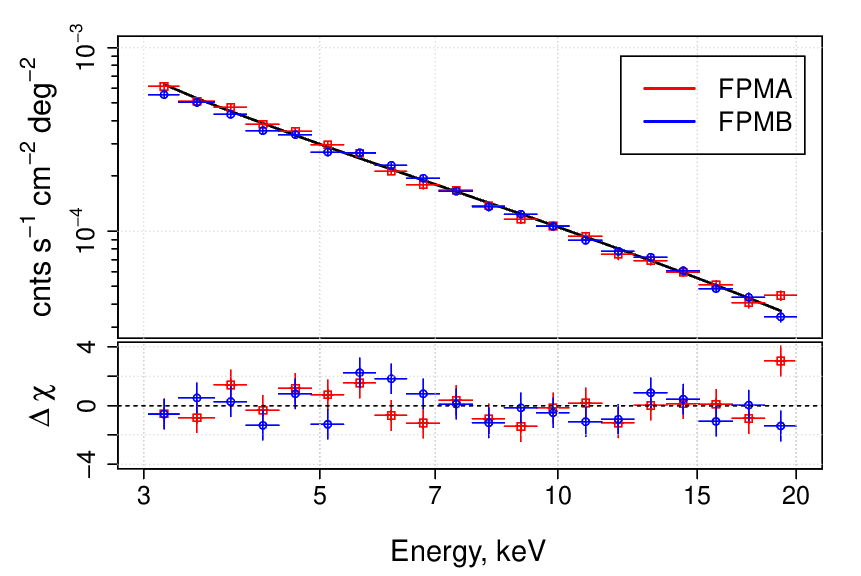}
\caption
   { \label{fig:cosmos:ep1:sci2} 
\nu\ spectrum of the aperture background component obtained from the COSMOS {\it EP1} field using the \sci\ data and corrected for the low-energy excess. The model is fitted in the 3--20~keV energy range.}
\end{figure}

As described in \cite{2014ApJ...792...48W}, the low-energy background component is related to solar photons reflected from the back of the aperture stop, and produces a time-variable signal dependent on whether or not the observatory is in sunlight. To the best of our knowledge, at least part of this scattered component, averaged over different spacecraft orientations, can follow a gradient similar to that produced by the aperture CXB (see also \S\ref{sect:backcxb}). In the current analysis, we detect a spatial low-energy component in \occ\ data and a strong excess in \sci\ data at the same energies. The question whether these components have the same origin needs further investigation, which is beyond the scope of this study and will be addressed in future work. For the current analysis, we fix the spectral shape of the \aocc\ component at $E<5$~keV and apply it to the spectrum of \asci\ with free normalization, similarly to the spectral analysis of the \nu\ background in \cite{2019ApJ...884..153P} and \cite{2020PhRvD.101j3011R}. Figure~\ref{fig:cosmos:ep1:sci2} shows the resulting spectrum of the \asci\ component for the COSMOS {\it EP1} data. Compared to Fig.~\ref{fig:cosmos:ep1:sci}, the spectrum demonstrates good agreement with the canonical CXB model (equation~\ref{eq:g99}), showing no strong soft excess in the residuals. The fit statistics is $\chi^{2}_{\rm r}$/dof =$1.2/36$ compared to $\chi^{2}_{\rm r}$/dof =$1.3/38$ without the low-energy component. 

\section{Results}
\label{sect:res}

\begin{table}
\caption{Measured 3--20~keV CXB flux and fit statistics for each data set. The flux is given in units of $10^{-11}$\ergscm. The uncertainties are quoted at the $90\%$ confidence level.}
\label{tab:resid}
\begin{tabular}{lcccccccc}
\hline
ID & \multicolumn{2}{c}{FPMA} & \multicolumn{2}{c}{FPMB} \\
 & $F_{\rm 3-20\ keV}^{\rm CXB}$ & $\chi^{2}_{\rm r}$/dof & $F_{\rm 3-20\ keV}^{\rm CXB}$ & $\chi^{2}_{\rm r}$/dof \\
\hline
1 & $2.82\pm0.06$ & 1.3/18 & $2.85\pm0.05$ & 1.2/18 \\
2 & $2.78\pm0.06$ & 0.8/18 & $2.89\pm0.05$ & 0.5/18  \\
3 & $2.89 \pm0.05$ & 0.7/18 & $2.85\pm0.04$ &  1.8/18 \\
4 & $2.68\pm0.05$& 1.2/18 & $2.70\pm0.04$ & 1.4/18  \\
5 & $2.81\pm0.05$ & 0.9/18 & $2.71\pm0.04$ &  0.7/18 \\
6 & $2.89\pm0.05$ & 1.1/18 & $2.89\pm0.04$ &  0.8/18 \\
\hline
\end{tabular}
\end{table}

\subsection{CXB flux in the 3--20~keV energy band}

\begin{figure}
  \includegraphics[width=0.99\columnwidth]{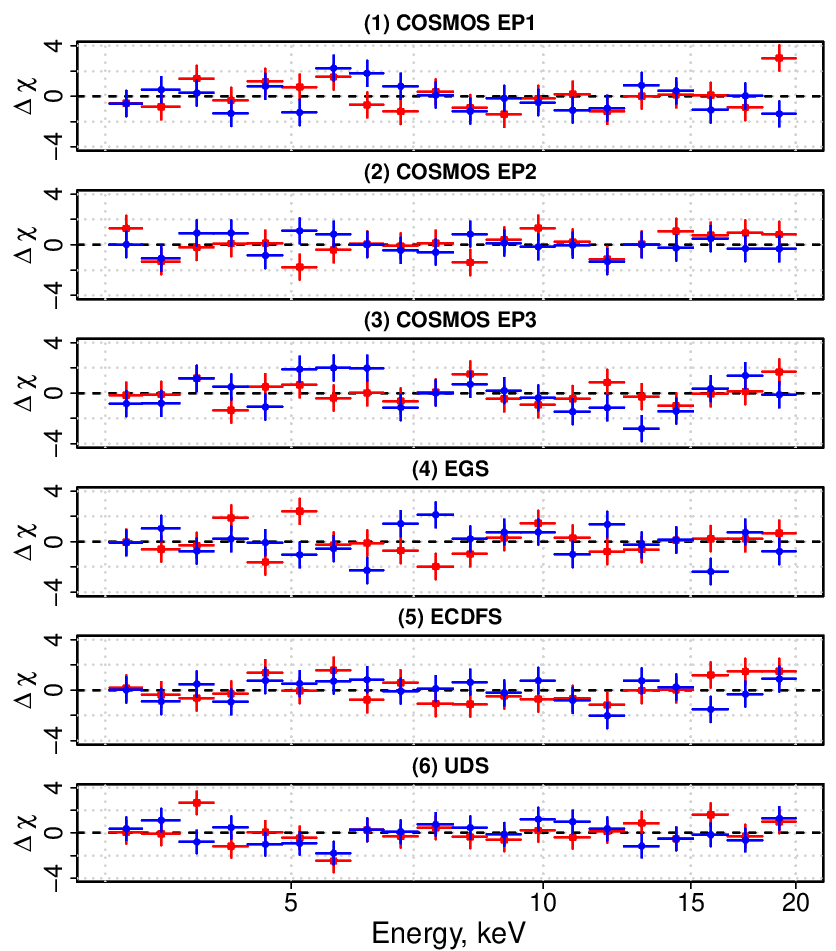}
\caption
   { \label{fig:resid} 
Spectral residuals of the aperture background component after subtracting the best-fit model consisting of the CXB component  (equation~\ref{eq:g99}) and the low-energy power-law component, both with free normalization.}
\end{figure}

We then applied spectral fitting with the model consisting of the CXB component (equation~\ref{eq:g99}) and a soft low-energy power-law component, both with free normalization as described above, to all the data sets. The normalization of the CXB component was estimated with the {\sc cflux} command in {\sc xspec} within the 3--20~keV energy range. As shown in Fig.~\ref{fig:resid}, the spectral fitting residuals do not strongly deviate from the best-fit model for any of the data sets. Table~\ref{tab:resid} presents the best-fit CXB normalization, errors at the $90\%$ confidence level and corresponding fit statistics. It can be seen that the CXB flux in the 3--20~keV band is measured at high significance $S/N=50-70\sigma$. The fit quality is good, except for data set \#3 when it reaches a marginally accepted level of $\chi^{2}_{\rm r}$/dof =$1.8/18$. 

\begin{figure}
  \includegraphics[width=0.99\columnwidth]{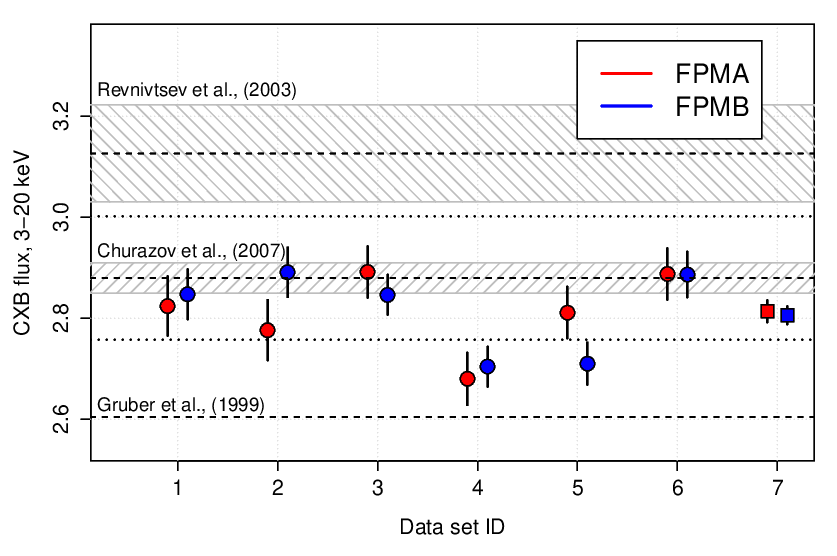}
\caption
   { \label{fig:cxb} 
\nu\ measurements of the CXB flux with FPMA (red) and FPMB (blue) in the 3--20~keV band for individual data sets 1--6 (Table~\ref{tab:data:num}). The square points at ID=7 show the weighted average over all the data sets. The flux is expressed in  units of $10^{-11}$~\ergscm. The dashed lines show the CXB 3--20~keV fluxes measured by \heao\ \citep{1999ApJ...520..124G}, {\it RXTE} \citep{2003A&A...411..329R}, and \int\ \citep{2007A&A...467..529C}. Note that the \rxte\ and \int\ measurements become consistent with each other if corrected to the same spectral model for the Crab Nebula. The shaded areas show the statistical uncertainties for the \rxte\ and \int\ measurements; the region between the dotted lines denotes the combined statistical and systematic uncertainty of the \int\ measurement \citep{2007A&A...467..529C}. }
\end{figure}

As shown in Fig.~\ref{fig:cxb}, the estimated CXB flux is largely consistent between independent FPMA and FPMB measurements, as it should be given that both modules observe related sky regions. This demonstrates high efficiency of the applied method for the two \nu\ modules with significantly different CXB spatial gradients. The different COSMOS epochs also show a consistent CXB flux, which further demonstrates reliability of the procedure. 

In Fig.~\ref{fig:cxb}, we also compare the \nu\ result with the canonical \heao\ \citep{1999ApJ...520..124G} result and with more recent CXB measurements by \rxte\ \citep{2003A&A...411..329R} and \int\ \citep{2007A&A...467..529C}. It is important to note that the \rxte\ result was obtained assuming a 8\% higher flux of the Crab nebula than adopted in the \int\ work, which directly translates into the normalization of the CXB flux \citep{2007A&A...467..529C}. After applying this scaling factor to the \rxte\ measurement shown in Fig.~\ref{fig:cxb}, it becomes consistent with the \int\ one within 1\%.

The \nu\ result is consistent with the \int\ measurement, $2.88\times10^{-11}$~\flux\ deg$^{-2}$, which is 10\% higher than the \heao\ one, $2.61\times10^{-11}$~\flux\ deg$^{-2}$. The average CXB flux over all the \nu\ data sets is $F_{\rm 3-20\,keV}^{\rm A}=(2.814\pm0.022)\times 10^{-11}$~\ergscm\ and $F_{\rm 3-20\,keV}^{\rm B}=(2.806\pm0.018)\times 10^{-11}$~\ergscm\ for FPMA and FPMB, respectively. The combined (FPMA and FPMB) \nu\ result for all the data sets combined and for the COSMOS field only is $F_{\rm 3-20\,keV}^{\rm AB, all}=(2.810\pm0.020)\times 10^{-11}$~\ergscm\ and $F_{\rm 3-20\,keV}^{\rm AB,COSMOS}=(2.860\pm 0.026)\times 10^{-11}$~\ergscm, respectively. 




%
%


\subsection{Flux variations and uncertainties}

\begin{figure}
  \includegraphics[width=0.99\columnwidth]{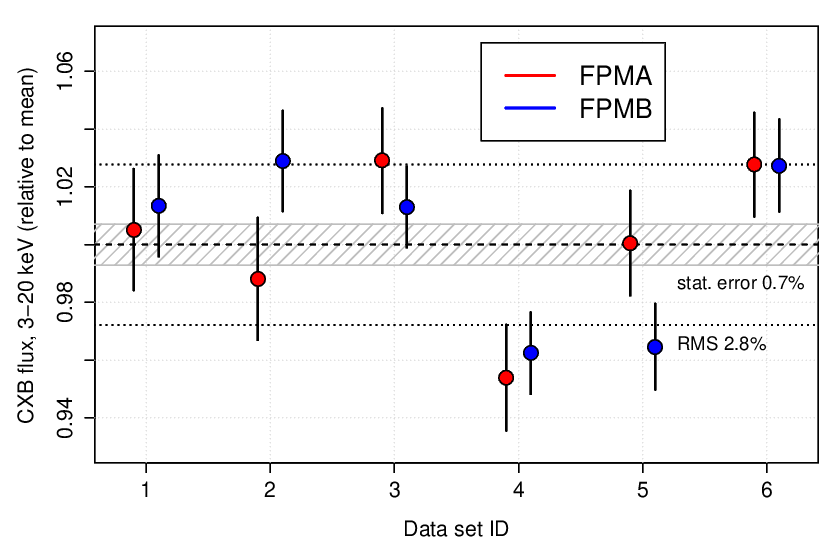}
\caption
   { \label{fig:cxb:ratio} 
Ratio of the CXB 3--20~keV flux measurements in the individual \nu\ data sets to the combined FPMA/FPMB mean $F_{\rm 3-20\,keV}^{\rm AB,all}=(2.81\pm0.02)\times 10^{-11}$~\ergscm. The shaded area denotes the statistical uncertainty of 0.7\% and the dotted lines show the RMS scatter of 2.8\%.}
\end{figure}

Figure~\ref{fig:cxb:ratio} shows the ratio of the \nu\ CXB flux measurements for the individual data sets to the mean flux $F_{\rm 3-20\,keV}^{\rm AB,all}$. The relative statistical 68\% uncertainty for the combined data set is 0.7\%. For comparison, the root-mean-square (RMS) scatter of the individual measurements is 2.8\%. The corresponding numbers for the COSMOS field only are 0.9\% and 1.6\%, respectively. Therefore, taking all the data sets into account, the statistical error is less than the RMS scatter by a factor of $\sim4$, whereas for the COSMOS field only, the statistical error is consistent with the data scatter within a factor of $\sim2$. This suggests that the 1.6\% RMS scatter in the COSMOS field is largely caused by statistical noise rather than by systematical uncertainties in the applied method, while the 2.8\% RMS scatter for the overall data set may be partially or predominantly associated with CXB variance, expected to play a role when dealing with observations in different directions over the sky. 

\subsubsection{Cosmic variance}

Since the CXB is the summed emission of unresolved extragalactic sources (AGNs), its intensity in a given direction is subject to Poisson variations in the number of sources, intrinsic source variability, and nearby large-scale structure \citep{1992ARA&A..30..429F}.  

To estimate the expected level of CXB variations between the different sky fields observed by \nu, we can use the \int\ extragalactic $\log N$--$\log S$ relation from \cite{2010A&A...523A..61K}, converted from 17--60~keV to 3--20~keV assuming a power-law spectrum with a slope of $1.8$. Following \cite{2008A&A...483..425R}, we estimate the relative uncertainty of the CXB flux in the 3--20~keV band as

\begin{equation}
\left(\frac{\delta I_{CXB}}{I_{CXB}} \right)_{\Omega}\sim 5.45\times 10^{-2} {S_{\rm max,11}^{1/4} \Omega_{\rm deg} ^{-1/2}},
\label{eq:poiss}
\end{equation}
where $S_{\rm max,11}$ is the maximum flux of undetected sources in units of $10^{-11}$~\flux, and $\Omega_{\rm deg}\approx5$ is the average solid angle of the \nu\ detector pixels. 

Based on our \nu\ measurements and that by \int\ \citep{2007A&A...467..529C}, we can adopt the mean CXB intensity to be $(2.8-2.9)\times10^{-11}$~\flux\ deg$^{-2}$. Using a conservative threshold of 1~mCrab\footnote{A flux of 1 mCrab in the 3--20~keV energy band corresponds to $2.5 \times 10^{-11}$~\flux\ for a source with a Crab-like spectrum.} for sources that are capable to produce a significant stray-light pattern on the \nu\ detectors, equation~(\ref{eq:poiss}) yields a CXB variance $\sim3\%$. 

We conclude that the expected CXB intensity variance for the \nu\ aperture is compatible with that actually measured by \nu\ between different sky fields (2.8\%).

\subsubsection{Geometric telescope model}

Our method makes use of the solid angle subtended by the open portion of the sky for a given pixel ($\Omega$ in equation~\ref{eq:prob}), which comes from the knowledge of the geometric model of the telescope, implemented in {\sc nuskybgd} code \citep{2014ApJ...792...48W}. The CAD (Computer-aided Design) model of the instrument, which includes the relative location of the optics bench, aperture stops, and detectors, was successfully used by many \nu\ stray-light studies of individual X-ray sources \citep{2014ApJ...781..107K,2015ApJ...814...94M,2017ApJ...841...56M,2021arXiv210201236G} and diffuse emission \citep[][]{2019ApJ...884..153P}. However, the geometric model is not fully calibrated with respect to the relative motion of the optical bench and the focal plane because of the non-rigid mast. The motion of the mast causes a floating shadow of the optical bench on the detector plane. The current analysis assumes a fixed mast of the telescope, which should be largely correct for long exposures on the average. We expect that measured CXB fluxes may change at a per cent level after the motion of the telescope's mast will have been fully calibrated \citep[see e.g.,][]{2016SPIE.9910E..0ZF}.

\subsubsection{Backscattered CXB emission}
\label{sect:backcxb}

As shown in \S\ref{sect:model}, the spectrum of the aperture component seen in the Earth occulted data (Fig.~\ref{fig:cosmos:ep1:occ}) is well approximated by a broken power law with $E_{\rm br}\simeq5$~keV. The low-energy ($E<5$~keV) component with steep $\Gamma_{1}\simeq5-7$ is related to the solar emission reflected from the back of the aperture stops, whose spatial gradient averaged over different telescope orientations to the Sun is similar to that produced by the aperture CXB. Apart from the soft solar component, the Earth occulted data also show a CXB-like component at $E>5$~keV, but with a harder spectral slope of $\Gamma_{2}\simeq1$. This additional high-energy component can be partially attributed to CXB emission reflected from the aperture stops (hereafter refereed to as the ``backscattered'' CXB). If such a backscattered CXB component were also present in direct (on-sky) CXB measurements, it could introduce a positive bias up to $\sim$10\%, estimated as the ratio of the 10--20~keV flux of the occulted spectrum (Fig.~\ref{fig:cosmos:ep1:occ}) to the 10--20~keV flux of the on-sky CXB aperture (Fig.~\ref{fig:cosmos:ep1:sci}). The fact that direct aperture and backscattered CXB emission produce similar detector gradients  makes it difficult to distinguish them from each other.

In this work we assume that backscattered CXB is effectively suppressed in on-sky observations, since Earth blocks the CXB from the back of the telescope. As proof that the backstattered CXB contribution has a minimal effect, \cite{2019ApJ...884..153P} presented a study of the Galactic Bulge Diffuse Emission (GBDE) using \nu's stray-light aperture. The authors measured the absolute 3--20~keV flux of the GBDE, which, if converted to luminosity per stellar mass, $L_{3-20\,keV}/M \approx (3.4 \pm 0.3) \times 10^{27}$\,erg\,s$^{-1}$\,$M_\odot^{-1}$, is in good agreement with the measurement by \rxte\ in the bulge and ridge, $L_{3-20\,keV}/M \approx (3.5 \pm 0.5) \times 10^{27}$\,erg\,s$^{-1}$\,$M_\odot^{-1}$~\citep{2006A&A...452..169R,2012MNRAS.424.2330R}. We conclude that the contribution of the backscattered CXB emission should not significantly bias our measurements. However, this question needs to be addressed in more detail, which will be done in future work.

\subsection{Average CXB spectrum}

\begin{figure}
  \includegraphics[width=0.99\columnwidth]{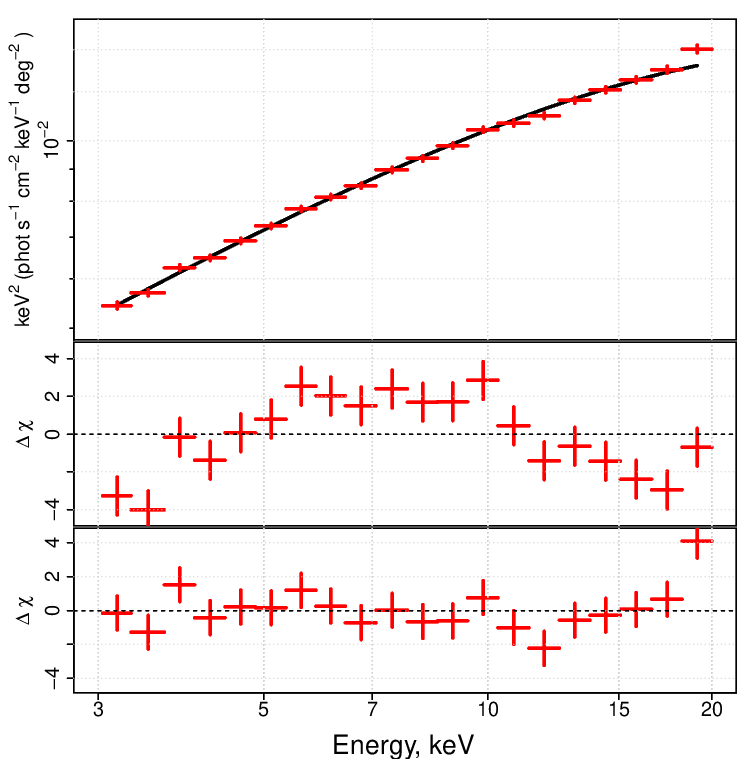}
\caption
   { \label{fig:cxb:final} 
Summed \nu\ CXB spectrum, averaged over all data sets and combining FPMA and FPMB. The middle and bottom panels show, respectively, the residuals for a power-law model with $\Gamma\simeq 1.5$ and for the canonical CXB spectral model (equation~\ref{eq:g99}) with free normalization.}
\end{figure}

Figure~\ref{fig:cxb:final} shows the average spectrum in the 3--20~keV energy band, obtained by stacking the CXB spectra obtained by both FPMA and FPMB for all the data sets. The low-energy component was subtracted from each individual spectrum before stacking. 

We first approximated the average spectrum by a simple power-law model, which gave bad fit statistics $\chi^{2}_{\rm r}$/dof =$4.5/18$ with $\Gamma\simeq1.5$. The residuals show a strong spectrum curvature, as seen from the second panel in Fig.~\ref{fig:cxb:final}. The fit by a canonical CXB spectral shape (equation~\ref{eq:g99}) provides much better quality $\chi^{2}_{\rm r}$/dof =$1.6/19$. The estimated 3--20~keV flux is $F_{\rm 3-20\,keV}=(2.81\pm 0.01)\times 10^{-11}$~\ergscm, consistent with our previous results presented in \S\ref{sect:res}. 

We conclude that the quality of the stacked 3--20~keV spectrum allows the spectral model to ``feel'' the high-energy rollover in the CXB spectrum at $\gtrsim30$~keV. To check this, we let the energy folding parameter (fixed at $E_{\rm fold}=41.13$~keV so far) in equation~(\ref{eq:g99}) to vary. This gave somewhat worse fit statistics $\chi^{2}_{\rm r}$/dof =$1.7/18$, however with $E_{\rm fold}=42.2\pm1.7$~keV compatible with \cite{1999ApJ...520..124G}. Note that the worsening of the fit is mainly caused by a single outlier data point at $E\simeq20$~keV, and limiting the fit to $E<18$~keV provides acceptable fit statistics $\chi^{2}_{\rm r}\simeq1.0$. Thawing in addition the $\Gamma$ parameter in equation~(\ref{eq:g99}) gives better fit quality $\chi^{2}_{\rm r}$/dof =$1.5/17$, however, at the price of a weaker constraint on the energy folding parameter: $E_{\rm fold}=54.4_{-9.9}^{+15.5}$~keV. This reflects the obvious fact that the data below 20~keV cannot well constrain the CXB spectral shape at higher energies.

\section{Conclusions}

We have demonstrated that using the well-known spatial modulation of the CXB flux on the \nu\ focal plane detectors, it is possible to measure the spectrum of the CXB in the 3--20~keV energy range. The presented method allows one to significantly detect the aperture CXB intensity up to 20~keV and distinguish it from instrumental background. The method is model-independent in the sense that the CXB intensity is estimated independently in different energy bins without using any a priori information. However, at energies $E<5$~keV we had to apply a correction for excess soft energy emission, calibrated on Earth occultation data, probably associated with the scattered emission from the Sun and producing a spatial gradient resembling that of the CXB.

Based on the \nu\ observations of four extragalactic fields, COSMOS, EGS, ECDFS and UDS, with a total exposure of 7~Ms, we estimated the CXB 3--20~keV flux at $(2.81\pm0.02)\times 10^{-11}$~\ergscm, which is $\sim8\%$ higher than measured with \heao\ \citep{1999ApJ...520..124G} but consistent with that measured by \int\ \citep{2007A&A...467..529C} and by \rxte\ (after correcting for the Crab flux normalization, \citealt{2003A&A...411..329R}). The inferred CXB spectral shape in the 3--20~keV energy band is fully consistent with the canonical model of \cite{1999ApJ...520..124G}. 

The RMS scatter of the CXB flux estimates for individual \nu\ data sets in the COSMOS field (1.6\%) is roughly compatible with the statistical errors (0.9\%) in this field, which indicates that the applied method is not strongly affected by systematic uncertainties and is limited mostly by pure statistical noise. The relative scatter 2.8\% of the CXB intensity in the different explored sky fields is compatible with the cosmic variance expected for the \nu\ stray-light aperture. This opens new possibilities for studying CXB anisotropy over the whole sky with the \nu\ data. For instance, CXB intensity accuracy at $\lesssim1\%$ level is required to detect the CXB variations related to mass concentrations in the nearby ($D<150$~Mpc) Universe, which allows to make an estimate of the total emissivity of low-luminosity AGNs \citep{2008A&A...483..425R}. It is also of interest to find the distortions and absorption features arising in the spectrum of the CXB radiation as it passes through the hot intergalactic gas in galaxy clusters \citep{2020AstL...45..791G}. Regarding future work, we plan to extend the current analysis onto the large all-sky ``serendipitous'' \nu\ data set with a total exposure of 20~Ms \citep{2013ApJ...773..125A,2017ApJ...836...99L}.

This work may be considered a pathfinder to an accurate measurement of the CXB surface brightness in the broader 6--70~keV energy range, as proposed for the MVN\footnote{\textit{Monitor Vsego Neba} is translated from Russian as all-sky monitor.} experiment \citep{2014AstL...40..667R}, currently being developed at the Space Research Institute (IKI) for implementation in the Russian segment of the International Space Station.

\section*{Data availability}
This work is based on the \nu\ data publicly available through the HEASARC Archive (\url{https://heasarc.gsfc.nasa.gov}).

\section*{Acknowledgements}

This work has made use of data from the \nu\ mission, a project led by the California Institute of Technology, managed by the Jet Propulsion Laboratory, and funded by the National Aeronautics and Space Administration. This research has made use of the \nu\ Data Analysis Software (NuSTARDAS) jointly developed by the ASI Science Data Center (ASDC, Italy) and the California Institute of Technology (USA). This research has made use of data and/or software provided by the High Energy Astrophysics Science Archive Research Center (HEASARC), which is a service of the Astrophysics Science Division at NASA/GSFC. R.K. and S.S. acknowledge support for this research from the Russian Science Foundation (grant 19-12-00396). D.W. and S.R. acknowledge support for this work from the NASA Astrophysics Data Analysis Program 80NSSC18K0686.




\bibliographystyle{mnras}
\bibliography{main}







\bsp	
\label{lastpage}
\end{document}